# On the Coupling of Hamilton's Principle and Thermodynamic Extremal Principles

Klaus Hackl[1], Jiří Svoboda[2],Franz Dieter Fischer[3],

[1]Lehrstuhl für Mechanik - Materialtheorie, Ruhr-Universität Bochum, D-44780 Bochum, Germany

[2]Institute of Physics of Materials, Academy of Sciences of the Czech Republic, CZ-61662 Brno, Czech Republic

[3]Institut für Mechanik, Montanuniversität Leoben, A-8700 Leoben, Austria

**Abstract**

*Extremal principles can generally be divided into two rather distinct classes. There are, on the one hand side, formulations based on the Lagrangian or Hamiltonian mechanics, respectively, dealing with time dependent problems, but essentially resting on conservation of energy and thus being not applicable to dissipative systems in a consistent way. On the other hand, there are formulations based essentially on maximizing the dissipation, working efficiently for the description of dissipative systems, but being not suitable for including inertia effects. Many attempts can be found in the literature to overcome this split into incompatible principles. However, essentially all of them possess an unnatural appearance. In this work, we suggest a solution to this dilemma resting on an additional assumption based on the thermodynamic driving forces involved. Applications to a simple dissipative structure and a material with varying mass demonstrate the capability of the proposed approach.*

## 1. Introduction

Extremal principles have been very successful in the past as applied to the formulation and solution of physical problems. These principles provide the foundation of many theoretical formulations, starting at classical rigid body dynamics, moving on to stability analysis, the description of dissipative processes, see e.g. Ortiz and Stainier [1], Petryk [2], Yang and Stainier and Ortiz [3], the study of material microstructures, Govindjee and Miehe [4], applications to chemical reactions and phase transformation, Basak and Levitas [5] and finally even quantum field theory. Moreover, several numerical methods including the Finite Element Method are based on extremal principles. However, despite all the successes, extremal principles are divided into two rather distinct classes. There are, on the one hand side, formulations based on the Lagrangian or Hamiltonian mechanics, respectively, dealing with time dependent problems, but essentially resting on conservation of energy and thus being not applicable to dissipative systems in a consistent way. On the other hand, there are formulations based essentially on maximizing dissipation, working efficiently for the description of dissipative systems, but not suitable for including inertia effects.

Many attempts can be found in the literature to overcome this split into incompatible principles, but essentially all of them possess an unnatural appearance. In this work, we suggest a solution to this dilemma resting on a simple additional assumption on the compatibility of thermodynamic driving forces.

One of the first works, dealing with the role of dissipation in a Lagrangian formulation, was published by Leech [6] in 1958. The according Lagrange function $L = T - V$, with $T$ being the kinetic energy and $V$ the potential energy, has been extended by the Rayleigh dissipation function R, for details see p.24 in [6] and the most recent publication [7].

Finally, the Lagrangian formulation with respect to the variables $x_i$ and $\dot{x}_i$ (as their time derivatives) reads as

$$-\frac{d}{dt}\left(\frac{\partial L}{\partial \dot{x}_i}\right) + \frac{\partial L}{\partial x_i} - \frac{\partial R}{\partial \dot{x}_i} = 0 \quad , \quad i = 1,...n \quad . \tag{1}$$

Formally, Eq. (1) can be considered as the application of the "classical" Hamilton's Principle (HP) to a system described by the contributions L and R.

However, Leech argued that such a description does not lead to a useful formulation in the Hamiltonian sense, since the energy of the system is not constant, see Leech [6], p.52. The relation in Eq. (1), however, has been the base for the variational formulation, later been "denominated" as "Modified Hamiltonian's Principle", (MHP), in Leech's book, Chapt. VI. Concluding, one can consider Eq. (1) as the basis for extending the "classical" Lagrange Equations" for nonconservative (i.e. "dissipative") systems. Consequently, also the term "Extended Hamilton's "principle" has been introduced.

Some corresponding books have been published in the last fifty years, dealing with variational methods for nonconservative (e.g. as dissipative processes), see e.g. Gyarmati [8], Vujanovic and Jones [9] and the book by Scholle [10]. The last one deals with a very applicable form for several constructions of "Lagrange densities".

Before discussion of further literature, it shall be mentioned that Hamilton's Principle is generally applied to a control volume moving with initially fixed material particles, i.e., the control volume is fixed in its initial configuration, but time-dependent in its current configuration, for details see [11] and later [12].

Several publications concerning the MHP, e.g. dealing with "a work done on the system by forces which cannot be derived form a potential" as [13], [14], [15] and for a "changing mass" [7], appeared in the same time period, when the a.m. books appeared, mostly dealing with dissipation. Concerning developing of multicomponent systems particularly Rosof [13], realizing the Onsager relations, shall be mentioned. Somewhat later Svoboda and Turek [16] derived with the MHP the diffusion-controlled evolution of closed multicomponent solid-state systems. Unifying procedures for reversible and irreversible processes were published by Kotowski [17] and somewhat later Anthony [18], particularly oriented to thermodynamics of complex fluxes.

The MHP has shown to be also an efficient tool for studying phase transformations, e.g. ferroelectric phase transitions in perovskite, see e.g. [19], and also in quantum many-body systems [20]. The MHP, often denoted as "Extended Hamilton's Principle", has still shown to be an efficient tool to describe rather "classical" (mass, friction etc.) motion processes, see the rather recent publications by Kim et al [21], [22] Wang and Wang [23], and concerning circuits by Biolek et al [24].

Finally, the "extended Hamilton principle (see the MHP below) has been recently emphasized as an efficient tool for dealing with coupled problems and a dissipative microstructure evolution in, see e.g. two recent publications by Said, [25] and Junker and Balzani [26].

**2. A motivating example**

As motivation, let us consider a toy model consisting of a point mass fixed on a mass-less disk of radius $r$. Moreover, there is a damper attached to the disk, see Fig. 1. Note that the damper stays in horizontal direction.

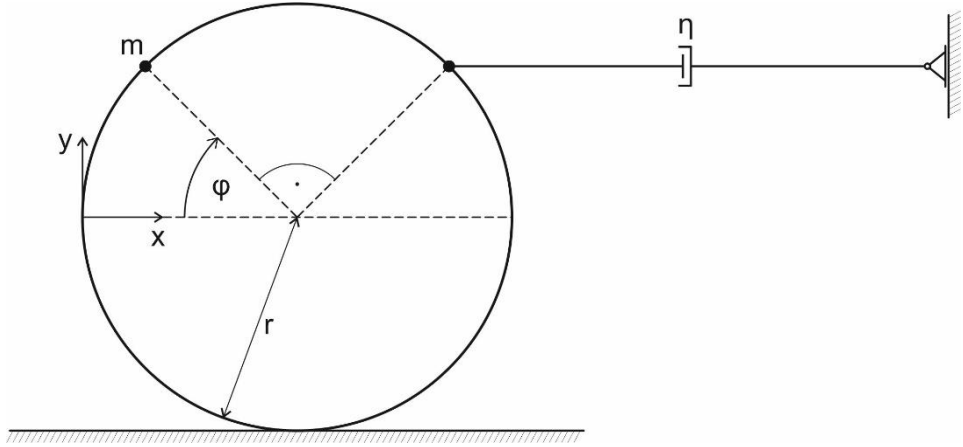

*Fig. 1: geometry of dynamic system with damper*

The kinematics of the system is described by a single generalized coordinate $q = \varphi$. For the damping force, we assume

$$F_d = \eta v_d = \eta r \dot{\varphi}(1 + \cos\varphi), \qquad (2)$$

where $v_d$ denotes the damper velocity and $\eta$ the damping constant. The system can be treated in a standard way by establishing a free body diagram and applying balance of linear and angular momentum leading to an equation of motion in the form

$$2r^2 m(1 + \sin\varphi)\ddot{\varphi} + r^2 m \cos\varphi \dot{\varphi}^2 + r^2 \eta (1 + \cos\varphi)^2 \dot{\varphi} + rmg\cos\varphi = 0. \qquad (3)$$

In the following, we would like to address the question, whether it is possible to obtain Eq. (3) in a consistent way employing a variational approach via a Thermodynamic Extremal Principle (TEP).

## 3. Extension of the Lagrangian

Let us introduce a Lagrangian, following the "derivation" by Hackl and Fischer [27] and further motivated by a recent analytical treatment of thermodynamics as in the recent publication [7]. The Lagrangian is formulated by Lagrangian coordinates (or may be denoted as general coordinates). According to [27] the internal variables $\mathbf{x}$ and their time derivatives $\dot{\mathbf{x}}$ are used and the Helmholtz free energy (or, better, Gibbs energy) as (contribution to) the potential energy $G(\mathbf{x})$. Consequently, it can already be referred to the according dissipation term, see Coleman and Gurtin [28]. Concerning more details, one may also read the recent publication by Fernándes and Quintavilla [29]. The dissipation function $Q$ follows as

$$Q(\mathbf{x},\dot{\mathbf{x}}) = \left(-\frac{\partial G}{\partial \mathbf{x}}\right):\dot{\mathbf{x}} = \mathbf{q}:\dot{\mathbf{x}} \geq 0 \ . \tag{4}$$

The vector $\mathbf{q}$ can be described as a "dissipative force" and will be dealt with below in context with the TEP, see Hackl and Fischer [27], Eq. (2.4), there as

$$\mathbf{q} = \frac{Q}{\partial Q/\partial \dot{\mathbf{x}} : \dot{\mathbf{x}}} \cdot \frac{\partial Q}{\partial \dot{\mathbf{x}}} \ . \tag{5}$$

Now we formulate the Lagrangian L with the Lagrangian coordinates $\mathbf{x}$, their time derivatives $\dot{\mathbf{x}}$, and time t as

$$L(\mathbf{x},\dot{\mathbf{x}},t) = K(\mathbf{x},\dot{\mathbf{x}}) - G(\mathbf{x},t) \tag{6}$$

with $K(\mathbf{x},\dot{\mathbf{x}})$ as kinetic energy. Here we want to mention that for the sake of some simplifications the contribution to L due to statistic aspects, see e.g. [30] on the Onsager-Machlup Function, is not included here.

The necessity of the current study is based on the still open question, if the TEP is in "agreement" with the Hamilton's Principle. Obviously, the Lagrangian formulation with the Rayleigh dissipation function R, see the Introduction, Eq. (1), as well as the pioneering paper on TEP by Svoboda and Turek [16] offer an according motivation, namely to extend the Lagrangian L in Eq. (6) to $\bar{L}$ by a term involving a generalized force $\mathbf{f}_d$ as

$$\bar{L}(\mathbf{x},\dot{\mathbf{x}},t) = L(\mathbf{x},\dot{\mathbf{x}},t) - \mathbf{f}_d \cdot \mathbf{x} \ . \tag{7}$$

Let us now perform the "Extended Hamilton's Principle" (MHP), to the above mentioned Lagrangian, see (7), yielding

$$-\frac{d}{dt}\frac{\partial \bar{L}}{\partial \dot{\mathbf{x}}} + \frac{\partial \bar{L}}{\partial \mathbf{x}} = -\frac{d}{dt}\frac{\partial K}{\partial \dot{\mathbf{x}}} + \frac{\partial K}{\partial \mathbf{x}} - \frac{\partial G}{\partial \mathbf{x}} - \mathbf{f}_d = \mathbf{0} \ . \tag{8}$$

Denoting the variational derivative of K in Eq. (8) by $\mathbf{D}_K = d/dt \ \partial K/\partial \dot{\mathbf{x}} - \partial K/\partial \mathbf{x}$, we can now interpret $\mathbf{f}_d$, see above Eq. (7), as an "extended" dissipation force term as

$$\mathbf{f}_d = -\mathbf{D}_K - \partial G/\partial \mathbf{x}. \tag{9}$$

As check, one may recognize that for $K \equiv 0$ the force term $\mathbf{f}_d$ becomes the "classical" dissipation force term $-\partial G/\partial \mathbf{x}$.

As starting quantity considered, the dissipation force $\mathbf{f}_d$ delivers in the product with the dissipation flux $\dot{\mathbf{x}}$ the dissipated energy D for a volume unit, which is identical with $Q(\dot{\mathbf{x}})$, see e.g. Eq. (4)

$$D = \mathbf{f}_d \cdot \dot{\mathbf{x}} = Q(\dot{\mathbf{x}}) = \mathbf{q} \cdot \dot{\mathbf{x}} \ . \tag{10}$$

The Thermodynamic Extremal Principle (TEP) is now based on the fact that Q or D obtains a maximum. This fact then allows the development of $\mathbf{q}$, see the derivation in [27], as

$$\mathbf{q} = \frac{Q}{\partial Q/\partial \dot{\mathbf{x}} : \dot{\mathbf{x}}} \cdot \frac{\partial Q}{\partial \dot{\mathbf{x}}} = \mathbf{f}_d \ . \tag{11}$$

The TEP originates from the seminal works by Ziegler et al [31, 32] and several later studies – for details we refer to [33]-[35].

The derivation from above can be found in [27], Eq. (2.4). Concluding, it follows with Eq. (9)

$$-D_K - \frac{\partial G}{\partial \mathbf{x}} = \frac{Q}{\partial Q/\partial \dot{\mathbf{x}} \cdot \dot{\mathbf{x}}} \cdot \frac{\partial Q}{\partial \dot{\mathbf{x}}} \ . \tag{12}$$

This Equation (12) delivers an implicit evolution equation for $\mathbf{x}(t)$.

## 4. Applications

**Example 1:** toy problem

We demonstrate that Eq. (3) can be obtained employing the TEP represented in Eq. (12). For this purpose, we have to establish the potentials K, G, and Q.

The kinetic energy K of the point mass is given, comparing Fig. 1, as

$$K = \frac{m}{2}(\dot{x}^2 + \dot{y}^2) = mr^2(1 + \sin\varphi)\,\dot{\varphi}^2 \ . \tag{13}$$

The potential energy G reads as

$$G = mgr\,\sin\varphi . \tag{14}$$

The dissipation is

$$D = F_d \cdot v_d = \eta v_d^2 = \eta r^2 (1+\cos\varphi)^2 \dot{\varphi}^2, \tag{15}$$

and Eq. (5) takes the form

$$q = f_d = F_d = \eta r (1+\cos\varphi) \, \dot{\varphi}, \tag{16}$$

compare Eq. (2). We obtain further

$$\frac{\partial G}{\partial \varphi} = mgr \, \cos\varphi, \tag{17}$$

and the variational derivative of K in Eq. (13) becomes

$$D_K = 2mr^2(1+\sin\varphi) \, \ddot{\varphi} + mr^2 \cos\varphi \, \dot{\varphi}^2. \tag{18}$$

Substitution of Eqs. (16), (17) and (18) into Eq. (12) yields Eq. (3) again, demonstrating that our approach provides the correct results.

**Example 2:** dynamics of a visco-elastic medium with variable density.

We consider a porous medium able to exchange mass with its environment, such that its density is dependent on its volume in a non-linear way as $\rho = \rho(\text{tr}\boldsymbol{\varepsilon})$. Note that we assume $\rho$ to be the density of the material in a fixed initial configuration. Thus, $\rho(\text{tr}\boldsymbol{\varepsilon})$ doesn't include the standard change of density due to elastic volume change, but only the additional density change due to mass exchange.

For simplicity, we neglect the elastic deformation, i.e. the Gibbs-energy satisfies $G \equiv 0$. Thus, the total deformation is an inelastic one, yielding

$$\boldsymbol{\varepsilon}_i = \boldsymbol{\varepsilon} = \frac{1}{2}(\text{grad } \mathbf{u} + (\text{grad } \mathbf{u})^T)$$

$$\tag{19}$$

with **u** as the displacement field.

Let us consider a control volume V. Then the total dissipation is given with $\boldsymbol{\sigma}$ denoting the stress state as

$$D_{tot} = \int_V \boldsymbol{\sigma} : \dot{\boldsymbol{\varepsilon}} \, dV. \tag{20}$$

We investigate a visco-elastic material of Norton-Hoff type by introducing a dissipation function with $\alpha > 0$ and $m > 0$ as material parameters,

$$Q(\dot{\boldsymbol{\varepsilon}}) = \alpha \|\dot{\boldsymbol{\varepsilon}}\|^m. \tag{21}$$

The constraint in Eq. (10) requires

$$\boldsymbol{\sigma} : \dot{\boldsymbol{\varepsilon}} = Q(\dot{\boldsymbol{\varepsilon}}). \tag{22}$$

Furthermore, applying Eq. (11) delivers with Eq. (21) the thermodynamic force $\mathbf{f}_d$ as

$$\mathbf{f}_d = \boldsymbol{\sigma} = \alpha \|\dot{\boldsymbol{\varepsilon}}\|^{m-2} \dot{\boldsymbol{\varepsilon}}. \tag{23}$$

The total kinetic energy follows with $\mathbf{v} = \dot{\mathbf{u}}$ and the density $\rho$ as function of $\mathrm{tr}\boldsymbol{\varepsilon}$ as

$$K_{tot} = \frac{1}{2} \int_V \rho(\mathrm{tr}\boldsymbol{\varepsilon}) v^2 dV. \tag{24}$$

Note the parallelism to Example 1, the kinetic energy being not only dependent on the rates, but on the variables themselves, here via the term $\mathrm{tr}\boldsymbol{\varepsilon}$. This gives rise to extra terms resulting from $\boldsymbol{D}_K$, which are very unusual in context of the mechanics of materials.

The Lagrangian in Eq. (7) reads now as

$$L = \int_{t_0}^{t_1} K_{tot} dt - \int_{t_0}^{t_1} \int_V \boldsymbol{\sigma} : \boldsymbol{\varepsilon} \, dVdt. \tag{25}$$

Variation of L with respect to $\mathbf{u}$ yields

$$\delta L = \int_{t_0}^{t_1} \int_V \rho(\mathrm{tr}\boldsymbol{\varepsilon}) \, \mathbf{v} \cdot \delta\mathbf{v} \, dVdt + \frac{1}{2} \int_{t_0}^{t_1} \int_V v^2 \rho'(\mathrm{tr}\boldsymbol{\varepsilon})(\mathrm{tr}\delta\boldsymbol{\varepsilon}) \, dVdt - \int_{t_0}^{t_1} \int_V \boldsymbol{\sigma} : \delta\boldsymbol{\varepsilon} \, dVdt = 0. \tag{26}$$

Partial integration of Eq. (26) with respect to space and time yields, ignoring the resulting boundary terms,

$$\delta L = -\int_{t_0}^{t_1} \int_V \frac{d}{dt} \rho(\mathrm{tr}\boldsymbol{\varepsilon}) \mathbf{v} \cdot \delta\mathbf{u} \, dVdt - \frac{1}{2} \int_{t_0}^{t_1} \int_V \mathrm{grad}\left[v^2 \rho'(\mathrm{tr}\boldsymbol{\varepsilon})\right] \cdot \delta\mathbf{u} \, dVdt$$
$$+ \int_{t_0}^{t_1} \int_V \mathrm{div}\, \boldsymbol{\sigma} \cdot \delta\mathbf{u} \, dVdt = 0 \tag{27}$$

for all $\delta\mathbf{u}$.

Equation (27) then implies

$$-\frac{d}{dt}\rho(\text{tr}\boldsymbol{\varepsilon})\mathbf{v} - \frac{1}{2}\text{grad}\left[v^2\rho'(\text{tr}\boldsymbol{\varepsilon})\right] + \text{div}\boldsymbol{\sigma} = 0. \tag{28}$$

Rewriting Eq. (28), using $\mathbf{v}=\dot{\mathbf{u}}$, yields

$$\text{div}\boldsymbol{\sigma} = \rho(\text{tr}\boldsymbol{\varepsilon})\ \ddot{\mathbf{u}} + \rho'(\text{tr}\boldsymbol{\varepsilon})\text{tr}\dot{\boldsymbol{\varepsilon}}\ \dot{\mathbf{u}} + \frac{1}{2}\text{grad}\left[\dot{u}^2\rho'(\text{tr}\boldsymbol{\varepsilon})\right]. \tag{29}$$

Interestingly, Eq. (29) displays two non-trivial terms in addition to the standard expression for the balance of linear momentum. This fact demonstrates nicely the importance and usefulness of the variational approach advocated in this work.

## 5. Conclusion

It is shown that the formulation of an "Extended Lagrangian" and application of the "Modified Hamilton Principle" (MHP) provide both, the thermodynamic forces and the according fluxes. Finally, coupling of the outcome of the MHP with the Thermodynamic Extremal Principle [TEP] delivers evolution equations for the internal variables. Neither "contradiction nor collision" of both a.m. principles can be observed. Examples show, that the approach developed here has the possibility of being applicable well beyond academic examples.

**References**

[1] Ortiz, M., Stainier, L., 1999. The variational formulation of viscoplastic constitutive updates, *Computer Methods in Applied Mechanics and Engineering*, **171**, Issues 3–4, 419-444.

[2] Petryk, H., 2003. Incremental energy minimization in dissipative solids, *Comptes Rendus Mécanique*, **331**, Issue 7, 469-474.

[3] Yang, Q., Stainier, L., Ortiz, M., 2006. A variational formulation of the coupled thermo-mechanical boundary-value problem for general dissipative solids, *J. Mech. Phys. Solids*, **54**, 401-424.

[4] Govindjee, S., Miehe, C., 2001. A multi-variant martensitic phase transformation model: formulation and numerical implementation, *Computer Methods in Applied Mechanics and Engineering*, **191**, 215-238.

[5] Basak, A., Levitas, V.I., 2018. Nanoscale multiphase phase field approach for stress- and temperature-induced martensitic phase transformations with interfacial stresses at finite strains, *J. Mech. Phys. Solids*, **113**, 162-196.

[6] Leech, J.W., 1958. Classical Mechanics (Methuen's Monographs on Physical Subjects), Methuen & Co. LTD, London.

[7] Podio-Guidugli, P., Virga, E. G., 2023. Analytical Thermodynamics, *J. Elasticity*, **153**, 787-812.

[8] Gyarmati, I., 1970. Non-Equilibrium Thermodynamics, (Field Theory and Variational Principles), Springer-Verlag, New York et al.

[9] Vujanovic, B.D., Jones, S.E., 1989. Variational Methods in Nonconservative Phenomena, Academic Press Inc., Boston et al.

[10]   Scholle, M., 1999. Das Hamiltonische Prinzip in der Kontinuumstheorie nichtdissipativer und dissipativer Systeme, Universität Paderborn.

[11]   McIver, D.B., 1973. Hamilton's principle for systems of changing mass, *J. Eng. Math.*, **7**, 249-261.

[12]   Steinboeck, A., Saxinger, M., Kugi, A., 2019. Hamilton's Principle for Material and Nonmaterial Control Volumes Using Lagrangian and Eulerian Description of Motion, *Appl. Mech. Rev.,* **71**, 010802, 14 pages.

[13]   Rosof, B.H., 1971. Hamilton's Principle and Nonequilibrium Thermodynamics, *Phys. Rev. A*, **4**, 1268-1274.

[14]   Kiehn, R.M., 1974. An extension of Hamilton's principle to include dissipative systems, *J. Math. Phys.,* **15**, 9-13.

[15]   Kaufman, A.N., 1984. Dissipative Hamiltonian systems: A unifying principle, *Phys. Letters,* **100A**, 419-422.

[16]   Svoboda, J., Turek, I., 1991. On diffusion-controlled evolution of closed-state thermodynamic systems at constant temperature and pressure, *Philos. Mag. B,* **64**, 749-759.

[17]   Kotowski, R., 1992. Hamilton's principle in thermodynamics, *Arch. Mech.*, **44**, 203-215.

[18]   Anthony, K.-H., 2001. Hamilton's principle and thermodynamics of irreversible processes – a unifying procedure for reversible and irreversible processes, *J. Non-Newtonian Fluid Mech.,* **96**, 291-339.

[19]   Paul, A., Sun, J., Perdew, J.P., Waghmare, U.V., 2017. Accuracy of first-principles interatomic interactions of ferroelectric phase transitions in perovskite oxides: Energy functional and effective Hamiltonian, *Phys. Rev. B,* **95**, 054111, 15 pages.

[20]   Beck, A., Goldstein, M., 2021. Disorder in dissipation-induced topological states: Evidence for a different type of localization transition, *Phys. Rev. B,* **103**, L241401, 7 pages.

[21]   Kim, J., 2012. Extended Hamilton's principle, *arXiv ,* arXiv:1204.0436, 31 pages.

[22]   Kim, J., Dargush, G.F., Ju, Y.-K., 2013. Extended framework Hamilton's principle for continuum dynamics, *Int. J. Solids Struct.,* **50**, 3418-3429.

[23] Wang, Q.A., Wang, R., 2018. A true least action principle for damped motion, *IOP Conf. Series: J. Physics*, **1113**, 012003, 5 pages.

[24] Biolek, Z., Biolek, D., Biolková, V., 2019. Hamilton's Principle for Circuits with Dissipative Elements, *Complexity,* **2019**, 2035324, 7 pages.

[25] Said, H., 2019. A Lagrangian-Hamiltonian unified formalism for a class of dissipative systems, *Math. Mech. Solids,* **24**, 1221-1240.

[26] Junker, P., Balzani, D., 2021. An extended Hamilton principle as unifying theory for coupled problems and dissipative microstructure evolution, *Cont. Mech. Thermodyn.,* **33**, 1931-1956.

[27] Hackl, K., Fischer, F.D. 2008. On the relation between the principle of maximum dissipation and inelastic evolution given by dissipation potentials, *Proc. R. Soc. A*, **464**, 117-132.

[28] Coleman, B. D., Gurtin, M., 1967. Thermodynamics with internal state variables, *J. Chem. Phys.*, **47**, 597-613.

[29] Fernandez, J. R., Quintanilla, R., 2023.On the hyperboli thermoelasticity with several dissipation mechanisms, *Arch. Appl. Mech.*, **93**, 2937-2945.

[30] Dürr D., Bach, A.m 1978. The Onsager-Machlup functions as Lagrangian for the most probable path of the diffusion process, *Commun. Math. Phys.*, **60**, 153-178.

[31] Ziegler, H., 1961. Zwei Extremalprinzipien der irreversibler Thermodynamik, *Ing. Arch.* **30** 410–416.

[32] Ziegler, H., Wehrli, C., 1987. On a principle of maximal rate of entropy production, *J.Non-Equilib. Thermodyn.* **12**, 229–243.

[33] Fischer, F.D., Svoboda, J., Petryk, H., 2014. Thermodynamic extremal principles for irreversible processes in materials science, *Acta Mater.* **67**, 1–20.

[34] Hackl, K., Fischer, F.D., Zickler, G.A., Svoboda, J., 2020. Are Onsager's reciprocal relations necessary to apply Thermodynamic Extremal Principles? *J. Mech. Phys. Solids* **135**, 103780, 6 pages.

[35] Hackl, K., Fischer, F.D., Svoboda, J., 2020. On the treatment of non-reciprocal rate independent kinetics via thermodynamic extremal principles, *J. Mech. Phys. Solids* **145**, 104149, 7 pages.